\documentclass[twocolumn,aps,pre,lengthcheck,showpacs]{revtex4}
\usepackage[dvips]{graphicx}
\usepackage{amssymb}
\usepackage{amsmath}
\usepackage{latexsym}
\usepackage{epsfig}
\usepackage{bm}
\usepackage{color}
\usepackage{times,psfrag,subfigure}      

\begin{document}
\title{Surveying the Free Energy Landscapes of Continuum Models:
\\ Application to Soft Matter Systems}
\author{Halim Kusumaatmaja}
\email{halim.kusumaatmaja@durham.ac.uk}
\affiliation{
Department of Physics, University of Durham, South Road, DH1 3LE, U.K.}
\date{\today}


\begin{abstract}
A variety of methods are developed for characterising the free energy landscapes of continuum, Landau-type free energy models. Using morphologies of lipid vesicles and a multistable liquid crystal device as examples, I show that the methods allow systematic study of not only the most relevant minimum energy configurations, but also the transition pathways between any two minima, as well as their corresponding energy barriers and transition state configurations. A global view of the free energy landscapes can then be visualized using either a disconnectivity graph or a network representation. Different forms of free energy functionals and boundary conditions can be readily implemented, thus allowing these tools to be utilised for a broad range of problems.
\end{abstract}
\maketitle

\section{Introduction}
In many soft matter systems the stability of minimum free energy configurations and the transition pathways from one state to another play a central role. Liquid crystal devices often exploit the existence of two or more (meta) stable states given the device geometry and surface anchoring~\cite{Spencer_2010,Tsakonas_2007,Majumdar_2012}, and are of great technological interests because power is only needed when the devices switch states. A major obstacle in many applications of superhydrophobic surfaces is the lack of stability of the suspended states~\cite{Kusumaatmaja_2008, Tuteja_2007}, where gas cushions are trapped in between the surface corrugations. Shape transformations of lipid bilayers hold the key to understanding important issues in cellular transport (e.g. budding phenomena, encapsulation of nanoparticles) as well as how cells react to external stimuli~\cite{Staykova_2011,Gompper_2014,Lipowsky_1993,Kusumaatmaja_2011}. All these examples require knowledge on the free energy landscapes, yet such calculations are usually limited to a certain facet of the landscapes, idealised scenarios, or both. Furthermore, while established methods exist to compute free energy barriers and transition states for atomistic simulations~\cite{Henkelman_climbing_2000,Henkelman_improved_2000,Trygubenko_2004,E_2002,Bolhuis_2002}, equivalent methods at the continuum level~\cite{Ren_2014,Ting_2011} are still lacking.

It is therefore my aim here to describe a suite of methods capable of fully characterising the free energy landscapes of continuum models, including basin-hopping global optimization~\cite{Li_1987,Wales_1997}, doubly-nudged elastic band~\cite{Trygubenko_2004}, and a hybrid eigenvector-following~\cite{Wales_book} techniques. These methods will allow us to calculate not only the minimum energy states, but also the transition state configurations, the height of the energy barriers, and the transition pathways between local minima. In some cases, the analysis will in fact show that there could be several competing pathways. It is also worth noting that, while these methods are well-known for atomistic systems, this paper takes the important step of reformulating them in a suitable form for continuum systems.

I will focus on cases where the physical systems of interest can be described using Landau theory. The free energy of the system is thus written in terms of expansion in scalar, vector and/or tensor fields. To highlight the strength and versatility of the methods, I will discuss examples from the field of soft matter: (i) shape transformations of lipid vesicles and (ii) multi-stable nematic liquid crystal devices. It is straightforward to implement other Landau free energy models which are routinely used in other areas of physics, chemistry and materials science. Complex boundary conditions (including Neumann, Dirichlet, periodic) can also be accounted while keeping the core of the simulation methodologies unchanged. 

This paper is organized as follows. The details of the computational methodologies and implementation are described in sections II and III. The methods are then used to study two different problems in soft matter and the results are presented in section IV. I summarise the most salient points of this paper in section V.

\section{The Free Energy Functionals}
Conceptually the most important step in the implementation of energy landscape methods is to define the energy (scalar) function to be optimized in terms of the degrees of freedom. In particle-based simulations, the degrees of freedom are usually the positions and orientations of the molecules. In field theories, they are the values of the fields at given coordinates in space. To illustrate how the free energy functionals can be defined in a suitable form for the methods described in the section III, I will now consider two specific examples. Here I shall assume it is appropriate to discretize space in a uniform grid, and for simplicity, I will choose a square grid in two dimensions and a cubic grid in three dimensions. 

\subsection{Phase Field Model for Lipid Vesicles}
To model the shape of lipid vesicles, I employ the phase field formulation proposed by Du and co-workers~\cite{Du_2004,Du_2008}. In this model, the total free energy is given by
\begin{eqnarray}
\Psi &=& \Psi_b + \frac{1}{2}k_V (V-V_0)^2 + \frac{1}{2} k_A (A-A_0)^2 .
\label{VesicleTotal}
\end{eqnarray}
The first term is the bending energy,
\begin{eqnarray}
\Psi_b &=& \int_V \frac{\kappa\epsilon}{2} \left( \Delta{}\phi - \frac{1}{\epsilon^2}(\phi^2-1)\phi \right)^2 dV,
\label{VesicleBending}
\end{eqnarray}
where a scalar order parameter $\phi(\bf{x})$ is used to distinguish the inside ($\phi = 1$) and outside ($\phi = -1$) of the vesicle, $\epsilon$ is the width of the interface, and $\kappa$ is the bending rigidity. The second and third terms in Eq.~\eqref{VesicleTotal} are penalty terms to conserve the vesicle volume and area. $V_0$ and $A_0$ are the target volume and area, and $k_V$ and $k_A$ are constants. I use the following definitions to compute the volume and area of the vesicle,
\begin{eqnarray}
V &=& \int_V \left(\frac{\phi+1}{2}\right) dV, \label{volume} \\
A &=& \sqrt{\frac{9}{8}} \int_V \left(\frac{1}{\epsilon} \left( -\tfrac{1}{2} \phi^2 + \tfrac{1}{4} \phi^4 \right) + \frac{\epsilon}{2} |\nabla\phi|^2\right) dV. \label{area}
\end{eqnarray}

The above free energy functional must be discretized and written in terms of the relevant degrees of freedom. The goal is to describe the free energy and its derivatives as functions of $\{\phi_{ijk}\}$, where the subscript $i (= 1 \dots N_x)$, $j (= 1 \dots N_y)$, and $k (= 1 \dots N_z)$ label the lattice points in three dimensions. The bending energy, vesicle volume and area can be expressed as
\begin{eqnarray}
\Psi_b &=& \sum_{ijk} \frac{\kappa\epsilon}{2}  \left( (\Delta{}\phi)_{ijk} - \frac{1}{\epsilon^2}(\phi_{ijk}^2-1)\phi_{ijk} \right)^2 \Delta{}x\Delta{}y\Delta{}z , \nonumber \\
V &=& \sum_{ijk} \left(\frac{\phi_{ijk}+1}{2}\right) \Delta{}x \Delta{}y \Delta{}z ,  \\
A &=& \sqrt{\frac{9}{8}} \,\, \sum_{ijk}  \frac{1}{\epsilon} \left( -\frac{1}{2} \phi_{ijk}^2 + \frac{1}{4} \phi_{ijk}^4 \right)  \Delta{}x \Delta{}y \Delta{}z \nonumber \\
&+& \sqrt{\frac{9}{8}} \,\, \sum_{ijk} \frac{\epsilon}{2} \left( (\partial_{x}\phi)_{ijk}^2 + (\partial_{y}\phi)_{ijk}^2 + (\partial_{z}\phi)_{ijk}^2 \right) \Delta{}x \Delta{}y \Delta{}z. \nonumber
\end{eqnarray}

Specific stencils are needed to approximate the derivatives. For the Laplacian, I use
\begin{equation}
(\Delta{}\phi)_{ijk} = (\partial^2_x \phi)_{ijk} + (\partial^2_y \phi)_{ijk} + (\partial^2_z \phi)_{ijk},
\end{equation}
with 
\begin{equation}
(\partial^2_x \phi)_{ijk} = \frac{\phi_{(i+1)jk} + \phi_{(i-1)jk} - 2 \phi_{ijk} }{(\Delta{}x)^2}, \nonumber 
\end{equation}
and similarly for the $y$ and $z$ directions. For the square-gradient term, one possibility is given by
\begin{equation}
(\partial_{x}\phi)_{ijk}^2  = \frac{\left[ (\phi_{(i+1)jk}-\phi_{ijk})^2 + (\phi_{(i-1)jk}-\phi_{ijk})^2 \right]}{2(\Delta{}x)^2}  , \label{stencilsq}
\end{equation}
and similarly for the derivatives in the $y$ and $z$ directions. Another possibility is to approximate $\partial_{x}\phi_{ijk} = (\phi_{(i+1)jk} - \phi_{(i-1)jk})/2$. However, this stencil is prone to a numerical checkerboard instability if the bending energy term is removed.  All the the stencils used here are second order accurate. 

{\it{Landau-de Gennes Free Energy}} - The liquid crystal application to be considered in this paper is a device first proposed by Tsakonas et al.~\cite{Tsakonas_2007}. The device consists of an array of square wells filled with nematic liquid crystals, and the well surfaces are treated such that the liquid crystal molecules lie parallel to the plane of the surfaces. This system is effectively two dimensional. Using the Landau-de Gennes framework~\cite{deGennes_book}, the liquid crystal configuration can thus be modeled by a symmetric traceless tensor,
\begin{equation}
{\bf{Q}} = \begin{bmatrix}
       Q_{11} & Q_{12}           \\
       Q_{12} & -Q_{11}
     \end{bmatrix}
= s(\bf{n}\otimes\bf{n}-I), \label{Qtensor}
\end{equation}
where $\bf{n}$ is the director and $s$ is a scalar order parameter that measures the degree of orientational ordering.

In the absence of external fields and surface effects, the Landau-de Gennes free energy under the one-constant approximation is given by~\cite{Majumdar_2012}
\begin{equation}
\Psi = \int_A \frac{\kappa_{el}}{2} |\nabla {\bf{Q}}|^2 dA + \int_A \left[-\alpha\mathrm{Tr}{\bf{Q}}^2+\frac{c^2}{4}(\mathrm{Tr}{\bf{Q}}^2)^2\right] dA. \label{feLC}
\end{equation}
The first integral corresponds to the elastic energy of the liquid crystals, and $\kappa_{el} > 0$ is an elastic constant. The second integral is the bulk free energy. The parameter $c$ is material-dependent constant. Given $T^*$ is the nematic-isotropic transition temperature, I define the coefficient $\alpha = \gamma (T^*-T)$, with $\gamma>0$ and $T$ is the temperature of the system. Here I shall assume $T < T^*$ such that $\alpha > 0$.

To reduce the number of input parameters in the model, I will now rewrite the free energy functional in its dimensionless form. Substituting Eq. \eqref{Qtensor} to \eqref{feLC} and defining $\tilde{x}=x/L$, $\tilde{\Psi} = c^2\Psi/\alpha^2L^2$, $\tilde{\kappa}_{el} = \kappa_{el}/\alpha L^2$, and $\tilde{\mathbf{Q}}^2 = c^2 \mathbf{Q}^2/\alpha$, I obtain
\begin{eqnarray}
\tilde{\Psi} &=& \int_{\tilde{A}} \tilde{\kappa}_{el} \left[|\tilde{\nabla} \tilde{Q}_{11}|^2 + |\tilde{\nabla} \tilde{Q}_{12}|^2\right] d\tilde{A}  \nonumber \\
&+& \int_{\tilde{A}} (\tilde{Q}_{11}^2+\tilde{Q}_{12}^2-1)^2  d\tilde{A}.  \label{feLCdim}
\end{eqnarray}
In this form, there is only one input parameter in the model, the dimensionless elastic constant $\tilde{\kappa}_{el}$. Note that $L$ is the size of the system, and from now on, I will drop the tilde in the equations. 

Using a square lattice, the discretized total free energy may be written as
\begin{eqnarray}
\Psi &=& \sum_{ij} \kappa_{el} \left[  |\nabla Q_{11}|_{ij}^2 + |\nabla Q_{12}|_{ij}^2\right] \Delta{}x\Delta{}y \nonumber \\
&+& \sum_{ij}  (Q_{11}^2+Q_{12}^2-1)_{ij}^2 \Delta{}x\Delta{}y  . 
\end{eqnarray}
Since the form of the gradient terms in Eq. \eqref{feLCdim} is the same as in \eqref{area}, the stencil described in Eq. \eqref{stencilsq} may be used. Thus, for either $Q_{11}$ and $Q_{12}$,
\begin{eqnarray}
|\nabla Q|_{ij}^2 &=& \frac{\left[ ( Q_{(i+1)j}-Q_{ij})^2 + (Q_{(i-1)j}-Q_{ij})^2 \right]}{2(\Delta{}x)^2}  \\
&+& \frac{\left[ ( Q_{i(j+1)}-Q_{ij})^2 + (Q_{i(j-1)}-Q_{ij})^2 \right]}{2(\Delta{}y)^2}. \nonumber
\end{eqnarray}

Suitable boundary conditions are needed at the solid walls. Here I simply consider the strong anchoring limit, where the director $\bf{n}$ lies parallel to the walls. In particular, I will follow the formulation given in Luo et al.~\cite{Majumdar_2012} where $(Q_{11}(x,y),Q_{12}(x,y))_{\mathrm{wall}} = s(x,y) (\cos 2\theta(x,y), \sin 2\theta(x,y))$, with $\theta(0,y)=\theta(1,y) = \pi/2$, $\theta(x,0)=\theta(x,1)=0$, and $s(t,0)= s(t,1)=s(0,t)=s(1,t) = f(t)$. I define
\begin{equation}
  f(t)   = 
\begin{cases}
    t/d ,& 0 \leq t \leq d, \\
    1,   & d \leq t \leq 1-d, \\
    (1-t)/d & 1-d \leq t \leq 1, 
\end{cases} 
\end{equation}
with $d$ chosen to be $3 \sqrt{\kappa_{el}}$. The strong anchoring boundary conditions are implemented as Dirichlet boundary conditions on the ${\bf{Q}}$ tensor, which can be applied via additional virtual lattice nodes along the walls. They influence the optimization of the free energy functional through the gradient terms.

\section{Energy Landscape Methodologies} 

\subsection{Finding Minimum Free Energy Configurations}

One of the primary goals in energy landscape methods is to characterise potential/free energy landscapes in terms of the minima and transition states (or more generally, saddle points). 

To obtain relevant local minimum configurations (low-lying energy states), I apply the basin-hopping algorithm~\cite{Li_1987,Wales_1997}. In basin-hopping, a step consists of a trial move followed by an energy minimization. The simplest trial move consists of random perturbations for the lattice field values, $\phi'_{ijk} = \phi_{ijk} + \xi d$, and similarly for the ${\bf{Q}}$-tensor. Here $\xi$ is a random number between -1.0 and 1.0, and $d$ is the amplitude of the perturbation, which I usually take to be $d = 0.125 \dots 0.5$. The energy minimization is carried out using the limited-memory Broyden-Fletcher-Goldfarb-Shanno (LBFGS) algorithm~\cite{Nocedal_1980,Liu_1989}. All that is required for the LBFGS algorithm is the gradient of the free energy with respect to the degrees of freedom, e.g. $d\Psi/d\phi_{ijk}$ for the phase field model of lipid vesicles.

The move is then accepted or rejected based upon the change in the corresponding free energy for the local minimum, $\Psi$. A simple approach, which has been widely used in the literature, is to use a Metropolis acceptance criterion. I further note that, for continuum systems, the number of minima is often very small ($<10$). In this case, it could be preferable to accept all steps, and as such, search for the majority, if not all, of the possible minima.
 
\subsection{Transition State Search} 

Methods for finding transition states (saddle points in the free energy landscapes) can generally be divided into single- and double-ended search methods. Here I first use a doubly-nudged elastic band (DNEB) method~\cite{Trygubenko_2004} to find candidate(s) for transition state(s) between every pair of minima. Note that any two pair of minima need not be connected by a single transition state. These candidates are then refined using a hybrid eigenvector-following technique~\cite{Wales_book}, which is a single-ended search technique. An alternative transition state finding method which has begun to find applications in continuum models is the string method~\cite{Ren_2014,Ting_2011}. 

Once the transition states are found, small displacements are applied in the two downhill directions. Energy minimizations are carried out using the LBFGS algorithm~\cite{Nocedal_1980,Liu_1989}. Such a run results in the minimum energy pathway for a minimum--transition state--minimum triplet. In some cases, new minimum energy states may be identified after this step. By iterating the process, I can systematically build a database of minima and transition states, and how they are interconnected.

{\it{Doubly-Nudged Elastic Band}} - The first step in the DNEB approach is to make an initial guess for a set of images, $\{\bf{\Gamma}^1, \bf{\Gamma}^2, ..., \bf{\Gamma}^N\}$ between two endpoints (minima) $\bf{\Gamma}^0$ and $\bf{\Gamma}^{(N+1)}$. I use the symbol $\bf{\Gamma}^\alpha$ as a shorthand for all the degrees of freedom in an image $\alpha$. For the vesicle application, it corresponds to $\{\phi_{ijk}\}^\alpha$, whereas for the liquid crsytal application, it refers to $\{(Q_{11},Q_{12})_{ijk}\}^\alpha$. I typically use 25 images between the two endpoints.

The simplest way to obtain an initial guess is to use linear interpolation of the lattice field distributions between the two minima (end points). While this is suitable for the liquid crystal application, my results indicate that such linear interpolation is inefficient for phase fields models. To illustrate this point, consider a liquid droplet (or a vesicle). Here the scalar field $\phi$ tends to two distinct bulk values, $+1$ inside the droplet and $-1$ outside, with a smooth transition at the interface over a finite width of several lattice spacings. If one linearly interpolates the phase field distributions between two shapes of droplets, there will be a large number of lattice points in the intermediate images, which are not at the interface, yet their phase field values deviate significantly from $+1$ and $-1$. These configurations are not physically relevant. Instead, one needs an algorithm that linearly interpolates the interfacial profile, and thus the droplet shape.

One possible algorithm is as follows. Generally, any lattice point in the simulation box falls into one of three categories. In the first category, the lattice point is inside the droplet/vesicle in both minima. I assign $\phi_{ijk} = 1$ for all the images. In the second category, the lattice point is always outside the droplet/vesicle and $\phi_{ijk} = -1$ can be assigned to all images. Further, the boundary profiles for these two categories may be computed. They are represented by two series of points, $\{ \bf{r}_i \}_{\mathrm{in}}$ and $\{ \bf{r}_i \}_{\mathrm{out}}$. In the third category, the lattice point is inside the droplet/vesicle in one minima and outside in the other minima. To assign a suitable value of $\phi_{ijk}$ in image $\alpha$, I compute the weighted distance squared
\begin{eqnarray}
\tilde{d}^2_{ijk,\mathrm{in}} = (1 - \tilde{\phi}^\alpha_{ijk})^2 d^2_{ijk,\mathrm{in}} , \nonumber \\
\tilde{d}^2_{ijk,\mathrm{out}} = (1 + \tilde{\phi}^\alpha_{ijk})^2 d^2_{ijk,\mathrm{out}} ,
\end{eqnarray}
where $\tilde{\phi}^\alpha_{ijk} = (1 - \tfrac{\alpha}{N+1}) \phi^0_{ijk} + (\tfrac{\alpha}{N+1}) \phi^{(N+1)}_{ijk} $. $d_{ijk,\mathrm{in}}$ and $d_{ijk,\mathrm{out}}$ are respectively the minimum distance between the corresponding lattice point and the two profiles, $\{ \bf{r}_i \}_{\mathrm{in}}$ and $\{ \bf{r}_i \}_{\mathrm{out}}$. If $\tilde{d}^2_{ijk,\mathrm{in}} < \tilde{d}^2_{ijk,\mathrm{out}}$, I set $\phi^\alpha_{ijk} = 1$. Otherwise, $\phi^\alpha_{ijk} = -1$. At this point, every image $\bf{\Gamma}^\alpha$ is a binary map of $\phi = \pm 1$. One can further smooth the image $\bf{\Gamma}^\alpha$ by taking advantage the fact that the scalar field follows a hyperbolic tangent profile near the interface~\cite{Du_2004}. 

The images are relaxed using energy gradients that include contributions from two components. The ``{\it{true}}'' gradient, ${\bf{g}^\alpha}$, depends on the derivatives of the the free energy functional with respect to the lattice degrees of freedom. The spring force component, $\widetilde{\bf{g}}^\alpha$, keeps the images roughly equidistant and is derived from the spring potential
\begin{eqnarray}
  V^\alpha = \frac{k}{2} \left((s^{\alpha,-})^2 - (s^{\alpha,+})^2\right)\,.
\end{eqnarray}
$s^{\alpha,-}$ and $s^{\alpha,+}$ are respectively the distances to the $\alpha-1$ (left) and $\alpha+1$ (right) images of image $\alpha$, $(s^{\alpha,\pm})^2 = \left| \bf{\Gamma}^{\alpha\pm1} - \bf{\Gamma}^\alpha \right|^2$. The coefficient $k$ may be taken as a constant. Alternatively, it can be dynamically adjusted based on the deviation of the image spacing from uniformity. 

Using the true and spring gradients without modification leads to corner-cutting (images are pulled away from the minimum energy path) and sliding-down problems (images slide down from barrier regions)~\cite{Henkelman_climbing_2000,Henkelman_improved_2000}. Certain projections for the gradients are needed based on the unit tangent $\bm{\hat \tau}^{\alpha}$ along the path. I follow the formulation in~\cite{Henkelman_improved_2000}, where $\bm{\hat \tau}^\alpha$ is defined based on the tangents to the left and to the right image, $\bm{\tau}^{\alpha,\pm} = \bf{\Gamma}^{\alpha\pm1} - \bf{\Gamma}^\alpha$. If the image $\alpha$ is at a maximum or a minimum, 
\begin{equation}
    \bm{\tau}^{\alpha} = 
\begin{cases}
    \bm{\tau}^{\alpha,+} \Delta \Psi^\alpha_{\mathrm{max}} + \bm{\tau}^{\alpha,-} \Delta \Psi^\alpha_{\mathrm{min}}  & \mathrm{if} \,\, \Psi^{\alpha+1} > \Psi^{\alpha} \\
    \bm{\tau}^{\alpha,+} \Delta \Psi^\alpha_{\mathrm{min}} + \bm{\tau}^{\alpha,-} \Delta \Psi^\alpha_{\mathrm{max}}  & \mathrm{if} \,\, \Psi^{\alpha+1} < \Psi^{\alpha} 
\end{cases}, 
\end{equation}
where
\begin{eqnarray}
\Delta\Psi^\alpha_{\mathrm{max}} = \mathrm{max}(|\Psi^{\alpha+1} - \Psi^{\alpha}|,|\Psi^{\alpha-1} - \Psi^{\alpha}|) ,\nonumber \\
\Delta\Psi^\alpha_{\mathrm{min}} = \mathrm{min}(|\Psi^{\alpha+1} - \Psi^{\alpha}|,|\Psi^{\alpha-1} - \Psi^{\alpha}|) .\nonumber
\end{eqnarray}
If the image $\alpha$ is neither at a maximum nor a minimum, then the tangent vector is defined as follows
\begin{equation}
    \bm{\tau}^{\alpha} = 
\begin{cases}
    \bm{\tau}^{\alpha,+}   & \mathrm{if} \,\, \Psi^{\alpha+1} > \Psi^{\alpha} > \Psi^{\alpha-1} \\
    \bm{\tau}^{\alpha,-}    & \mathrm{if} \,\, \Psi^{\alpha+1} < \Psi^{\alpha} < \Psi^{\alpha-1} 
\end{cases}.
\end{equation}
The unit tangent vector is defined as $\hat{\bm{\tau}}^\alpha={\bm{\tau}}^\alpha/|{\bm{\tau}}^\alpha|$.

The first projection in DNEB retains only the components of the true gradient that are perpendicular to the unit tangent vector $\hat{\bm{\tau}}^\alpha$,
\begin{align}
  \bf{g}^\alpha_\perp = \bf{g}^\alpha - \left( \bf{g}^\alpha \cdot \hat{\bm\tau}^\alpha \right) \hat{\bm\tau}^\alpha \, .
\end{align}
Removing the tangent part of the true gradient is numerically not sufficient to achieve equidistant images. An additional component from the spring force is required, and in DNEB, the following component of the spring constant gradient is retained~\cite{Trygubenko_2004},
\begin{eqnarray}
  \widetilde{\bf{g}}^\alpha_{\mathrm{DNEB}} &=& k \left( s^{\alpha,+} - s^{\alpha,-} \right) \hat{\bm\tau}^\alpha \nonumber \\
&+& \widetilde{\bf{g}}^\alpha_\perp - \left( \bf{\widetilde g}^\alpha_\perp \cdot \bf{\hat g}^\alpha_\perp \right) \bf{\hat g}^\alpha_\perp  \,,\label{eq:neb_spring}
\end{eqnarray}
where $\widetilde{\bf{g}}^\alpha_\perp = \widetilde{\bf{g}}^\alpha - \left(\widetilde{\bf{g}}^\alpha\cdot\hat{\bm\tau}^\alpha\right)\hat{\bm\tau}^\alpha$.

{\it{Eigenvector-Following Technique}} - To refine the transition states obtained in the doubly-nudged elastic band procedure, I use an eigenvector-following technique~\cite{Crippen_1971,Munro_1999,Wales_book}. In this work, rather than computing the complete Hessian (second derivatives of the free energy functional with respect to the lattice degrees of freedom), I use a Rayleigh-Ritz approach~\cite{Hildebrand} to estimate the smallest eigenvalue and the corresponding eigenvector. Since it only requires the gradient information~\cite{Munro_1999,Wales_book}, it is more efficient for large systems.

It is also essential that the zero eigenmodes are removed. This can be achieved using projection operators~\cite{Baker_1991,Wales_1993} or eigenvalue shifting~\cite{Wales_book}. In the models described here, the zero eigenmodes correspond to translations in lattice directions where periodic boundary conditions are applied. For a scalar field, $\{\phi_{ijk}\}$, and assuming periodic boundary condition in the $\alpha$-direction, the zero Hessian eigenvector is given by
\begin{equation}
{\bf{\hat{e}}} = \frac{1}{N} \left( \frac{\partial\phi_{111}}{\partial\alpha} \dots  \frac{\partial\phi_{ijk}}{\partial\alpha} \dots  \frac{\partial\phi_{N_xN_yN_z}}{\partial\alpha} \right) , \label{zeroeigen}
\end{equation}
where $N$ is a normalisation constant, such that $|{\bf{\hat{e}}}| = 1$. Periodic boundary conditions are not used for the liquid crystal application considered here. In cases where periodic boundary conditions are needed, the zero Hessian eigenvector can be obtained as in Eq. \eqref{zeroeigen}, with the scalar field $\phi$ replaced by the $\mathbf{Q}$-tensor.

\section{Results}

\subsection{Vesicle Morphologies} 

In this sub-section, I will consider the morphological transitions of single-component lipid vesicles. I shall assume that the spontaneous curvature of the membrane and the Gaussian bending modulus are zero. Thus, the shapes of the vesicle depend only on the reduced volume~\cite{Seifert_1991}, $\nu = \frac{V_0 / (4\pi/3)}{(A_0 / 4\pi)^{3/2}}$, where $V_0$ and $A_0$ are the volume and area of the vesicles. $\nu$ is tuned in the simulations by varying $A_0$, while keeping the volume $V_0$ constant. Here I have used  a $140 \times 140 \times 210$ grid and the following simulation parameters: $\kappa = 1.0$,  $\epsilon = 1.5 \Delta {}x = 0.03$, $k_A = k_V = 10^5$, and $V_0  = 1.375 \times 10^5 \Delta{}x^3$. I will also restrict the analysis to axisymmetric vesicles, though the methods themselves are not limited by this symmetry.

\begin{figure}
\centering
\includegraphics[scale=1.0,angle=0]{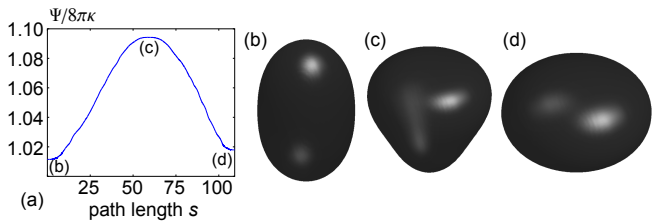}
\caption{Shape transition from a prolate (b) to an oblate (d) for a vesicle with reduced volume $\nu = 0.987$. Panel (a) shows the energy profile for the minimum energy path. The pear-shaped configuration in panel (c) is the corresponding transition state.}
\label{FigVesicle1}
\end{figure}
While the shapes of lipid vesicles have been analysed in much detail in the literature (see e.g. \cite{Du_2004,Du_2008,Seifert_1991,Seifert_1997}), these studies are usually limited to minimum energy configurations. My aim here is to show that, using the methods developed here, it is possible to systematically study not only the minima, but also the transition state configurations and the minimum energy paths between various vesicle morphologies. To illustrate this, I first consider a vesicle with a reduced volume of $\nu = 0.987$. When $\nu$ is close to 1, there are only two possible minimum energy shapes, a prolate (Fig. \ref{FigVesicle1}(b)) and an oblate (Fig. \ref{FigVesicle1}(d)). Furthermore, I find that there is only one transition state, the morphology of which is given in Fig. \ref{FigVesicle1}(c). Thus, the free energy landscape is very simple. Fig. \ref{FigVesicle1}(a) shows the minimum energy path for the transition from a prolate to an oblate; it corresponds to the transition pathway with the smallest possible energy barrier. The $x$-axis is a measure of distance in the configurational space between morphology $\alpha$ along the path to a reference minimum configuration, $(s^\alpha)^2 = \sum_{i=0}^\alpha \left| \{\phi_{ijk}\}^{i+1} - \{\phi_{ijk}\}^{i} \right|^2$. A movie showing the full transition pathway is available in the supporting information~\cite{SI}.

When the reduced volume differs significantly from 1, the free energy landscape becomes more complex. For $\nu = 0.661$, there are (at least) ten minimum energy shapes. These minima are obtained from a basin-hopping run of 500 steps. Solutions arising due to the periodicity of the grid are ignored. Doubly-nudged elastic band~\cite{Trygubenko_2004} and hybrid eigenvector-following methods~\cite{Wales_book} are then used to compute the transition states and minimum energy paths for transitions between these various morphologies.
\begin{figure}
\centering
\includegraphics[scale=0.625,angle=0]{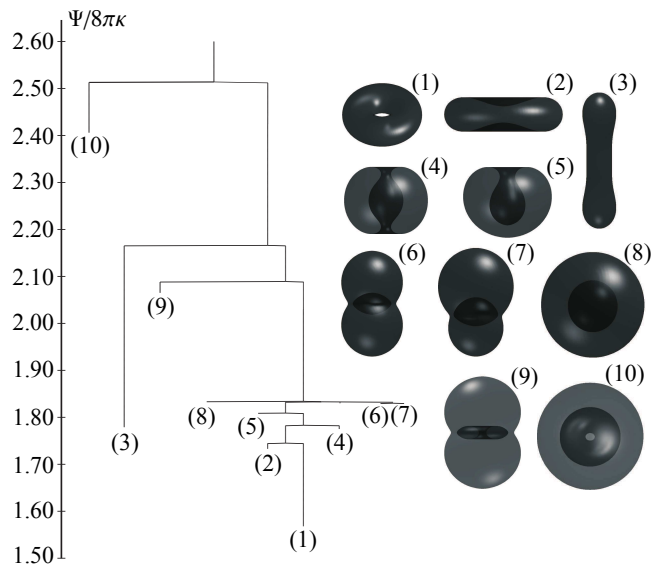}
\caption{A disconnectivity graph for a vesicle with reduced volume $\nu = 0.661$.}
\label{FigVesicle2}
\end{figure}

With an increasing number of minima and transition states, one needs a better way to visualize the free energy landscape. Plotting the energy profile as in Fig. \ref{FigVesicle1}(a) provides detailed information on the transition between any two minima. However, the ``{\it{global view}}'' of the landscape is not captured. One possible representation is using the disconnectivity graph~\cite{Becker_2003}, as shown in Fig. \ref{FigVesicle2}. Here the $y$-axis corresponds to the energy scale. The end of every line corresponds to a minimum. If one traces two minima, they meet at a vertex. This vertex corresponds to the highest energy along the minimum energy path for the transition between the two minimum configurations. Thus, a disconnectivity graph provides information on the competing minima and how easy for these minima to interconvert.

I note that the minimum energy shapes shown in Fig. \ref{FigVesicle2} have different genera (``{\it{number of holes}}''). Thus, in general, the Gaussian curvature term should be taken into account when the transition pathways are computed. Extensions of the model described in section IIA to include Gaussian curvature, multicomponent membrane and adhesion to solid substrate are available in the literature (see e.g. ~\cite{Du_2008,Lowengrub_2009,Das_2008}), and can be readily implemented in the energy landscape methods described here. These are avenues for future research and are outside the scope of this report. 

\subsection{Multistable Nematic Liquid Crystal Devices} 

\begin{figure}
\centering
\includegraphics[scale=1.0,angle=0]{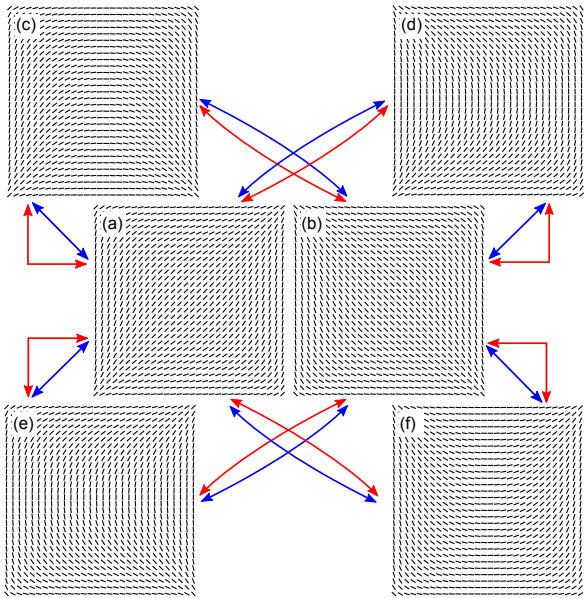}
\caption{Minimum energy configurations in a square liquid crystal device. There are two classes of configurations: (a-b) diagonal and (c-f) rotated states. Each double arrow corresponds to a minimum-transition state-minimum triplet, thus visualising the connectivity of the free energy landscape.}
\label{FigLC}
\end{figure}

The second application to be considered is a multistable nematic liquid crystal device~\cite{Tsakonas_2007,Majumdar_2012}. The device consists of an array of square wells filled with nematic liquid crystals, and the well surfaces are treated such that the liquid crystal molecules lie parallel to the place of the surfaces. One such cell is modeled in Fig. \ref{FigLC} using the framework of the Landau-de Gennes free energy, assuming strong anchoring on the surface. I use a $150 \times 150$ grid, and the dimensionless elastic coefficient is taken to be $\kappa_{el} = 4 \times 10^{-4}$.

Within such a well, there are two classes of minimum configurations, namely diagonal and rotated states~\cite{Tsakonas_2007,Majumdar_2012,Aarts_2014}. Due to the symmetry of the system, there are two equivalent diagonal states, see Fig. \ref{FigLC}(a-b), and four equivalent rotated states, see Fig. \ref{FigLC}(c-f). Since there are six minimum free energy states, in principle there are 15 transition pathways to be considered. However, transitions between any rotated and diagonal states are related by symmetry. Further, I find that the minimum energy path between any two rotated states or two diagonal states is composed by a sequence of rotated-diagonal transitions. A direct pathway between pairs of diagonal states or pairs of rotated states is never observed.  Thus, the free energy landscape of the system may be represented by a network shown in Fig.~\ref{FigLC}. Each line corresponds to a rotated state--transition state--diagonal state triplet. Such a network representation is an alternative to the disconnectivity graph used in the previous sub-section to visualise the landscape globally.

\begin{figure*}
\centering
\includegraphics[scale=1.1,angle=0]{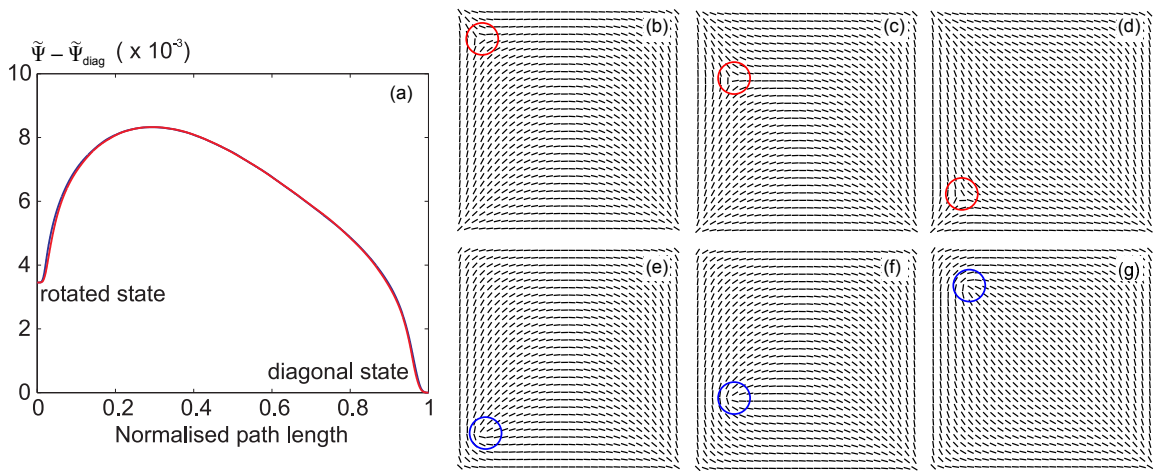}
\caption{Two competing pathways for a transition from a rotated to a diagonal state. The two mechanisms are respectively driven by a $-1/2$ (panel b-d) and a $+1/2$ (panel e-g) defect. The energy profiles along these two pathways are identical (panel a). The configurations in panels (c) and (f) correspond to the transition states.}
\label{FigLC2}
\end{figure*}

For a given rotated-diagonal pair, there are in fact two competing transition mechanisms. As shown in Fig. \ref{FigLC2}(a), the energy profiles for the two mechanisms are virtually the same. In the first mechanism, Fig. \ref{FigLC2}(b-d), a bend defect transforms to a splay defect in the top left corner. This results in an excess $-1/2$ defect which propagates to the bottom left corner, and subsequently accommodates a splay to bend defect transformation there. In the second mechanism, Fig. \ref{FigLC2}(e-g), the transition is driven by a $+1/2$ defect, which propagates from the bottom left to the top left corner.  The full pathways for these two mechanisms are available as movies in the supporting information~\cite{SI}. It is worth mentioning that the transition states with defects shown here are reminiscent of recent experimental observations in shallow nematic chambers \cite{Aarts_2014}. Higher energy pathways are also possible and are observed nimerically, corresponding to transition mechanisms mediated by two or more defects. This will be discussed in a separate publication.

\section{Discussions} 

Here I have described a set of methods suitable for mapping the free energy landscapes of continuum, Landau-type free energy models in terms of their minima, transition states, and minimum energy paths. Compared to studies which focus only on the minimum energy configurations, this approach provides several significant advantages. It allows us to systematically study the energy barriers and transition pathways as a function of the system parameters. On a fundamental level, it enriches any study which involves the construction of a phase or morphology diagram. As shown above, not only there are competing minima, but also competing pathways may exist for the transition between two minimum configurations. It is also possible to analyze the (dis)appearance of minima, and correspondingly, how the connectivity of the landscapes is altered as a result of changing the system parameters. On a more practical level, these energy landscapes techniques have great potential to be used as a design or optimization tool. Taking the multistable liquid crystal devices as an example, the minimum energy path corresponds to the most efficient transition pathway, including the minimum energy required to induce the transition between any two minimum states. It can also provide insights into the effective form of external perturbations (e.g. strength and direction of the external electric field) required for the transition. Additionally, since the configurations of the most relevant minima and transition states are known, these results may be exploited to design suitable surface treatments (anchoring strength, surface corrugations, etc), which increase/decrease the stability of the minima and alter the heights of the energy barrier for the transitions.

It is worth noting that it is straightforward to implement other forms of free energy functionals. Given the widespread use of Landau free energies in physics, chemistry, and materials science, these methods may therefore have wide-ranging applications. The descriptions and discretizations of the the free energy functionals are also compatible with other simulation techniques, including the implementation of complex boundary conditions, such as surface corrugations, curved and non-symmetric boundaries. I envisage the use of these energy landscape techniques to be highly complementary to methods which capture the dynamics of the system (such as finite element, lattice Boltzmann method, etc). 


{\bf{Acknowledgements}} - I thank Prof. D. J. Wales for useful discussions. I also acknowledge EPSRC (EP/J017566/1) for funding.

\bibliographystyle{aipnum4-1}
\bibliography{FELandscape}

\end{document}